# Thermally enhanced photoluminescence and temperature sensing properties of $Sc_2W_3O_{12}$:$Eu^{3+}$ phosphors

Yude Niu, Yuzhen Wang, Kaiming Zhu, Wanggui Ye, Zhe Feng, Hui Liu, Xin Yi, Yihuan Wang, Xuanyi Yuan[*]

Beijing Key Laboratory of Optoelectronic Functional Materials & Micro-nano Devices, Department of Physics, Renmin University of China, Beijing 100872, China..

Corresponding author:

Xuanyi Yuan: E-mail address: yuanxuanyi@ruc.edu.cn; telephone number: 0086-10-82501673; fax number: 0086-10-62517887; postal address: Renmin University of China, No. 59 Zhongguancun Street, Haidian District, Beijing 100872 P.R.China

**Abstract**

Currently, lanthanide ions doped luminescence materials applying as optical thermometers have arose much concern. Basing on the different responses of two emissions to temperature, the fluorescence intensity ratio (FIR) technique can be executed and further estimate the sensitivities to assess the optical thermometry performances. In this study, we introduce different doping concentrations of $Eu^{3+}$ ions into negative expansion material $Sc_2W_3O_{12}$, accessing to the thermal enhanced luminescence from 373 to 548 K, and investigate the temperature sensing properties in detail. All samples exhibit good thermally enhanced luminescence behavior. The emission intensity of $Sc_2W_3O_{12}$: 6 mol% $Eu^{3+}$ phosphors reaches at 147.81% of initial intensity at 473 K. As the Eu doping concentration increases, the resistance of the samples to thermal quenching decreases. The FIR technique based on the transitions $^5D_0 \rightarrow {}^7F_1$ (592 nm) and $^5D_0 \rightarrow {}^7F_2$ (613 nm) of $Eu^{3+}$ ions demonstrate a maximum relative temperature sensitivity of 3.063% $K^{-1}$ at 298 K for $Sc_2W_3O_{12}$: 6 mol% $Eu^{3+}$ phosphors. The sensitivity of sample decreases with the increase of $Eu^{3+}$ concentration. Benefiting from the thermal enhanced luminescence performance and good temperature sensing properties, the $Sc_2W_3O_{12}$: $Eu^{3+}$ phosphors can be applies as optical thermometers.

**Keywords:** photoluminescence, $Sc_2W_3O_{12}$: $Eu^{3+}$, negative lattice expansion, thermal enhanced luminescence,

## 1. Introduction

Currently, lanthanide ions doped luminescence phosphors have been widely applied in displays,[1] lighting,[2] and sensors[3]. Therein, optical thermometry based on luminescence materials have arose many attentions due to the fast response, non-contact and high sensitivity.[4, 5] By employing the fluorescence intensity ratio (FIR) of thermally coupled energy levels (TCELs) or non-thermally coupled energy levels (NTCELs), the temperature sensing properties can be obtained.[6, 7] The potential of luminescence materials for applied as optical thermometers is evaluated by absolute sensitivity ($S_a$) and relative sensitivity ($S_r$).[7, 8] For $Eu^{3+}$-doped materials, several pairs of emissions have been executed for FIR technique due to the abundant energy levels of $Eu^{3+}$ ions.[9, 10] Liang et al. revealed that $^5D_1/^5D_0\rightarrow^7F_1$ TCELs (541 nm/590 nm) and $(^5D_0\rightarrow^7F_2)/(^5D_1\rightarrow^7F_1)$ NTCELs (625 nm/541 nm) of $Eu^{3+}$ ions in $LiNbO_3$ single crystals can be devoted to the optical temperature sensor based on FIR method, and the maximum $S_a$ values are $7\times10^{-4}$ $K^{-1}$ and $24\times10^{-4}$ $K^{-1}$, respectively.[11] In addition, Nikolić et al. developed $(^5D_1\rightarrow^7F_1)/(^5D_0\rightarrow^7F_2)$ NTCELs (533 nm/611 nm) in $Gd_2O_3$:Eu phosphors for thermography, and the maximum $S_a$ is $7\times10^{-4}$ $K^{-1}$ at 800 K.[12]

Generally, the performance of luminescence materials for applications, including optical thermometry, is usually restricted by inevitable debasement of luminescence intensities with elevated temperatures, namely thermal quenching.[13] So far, many efforts have been devoted to resist thermal quenching.[14-16] Negative thermal expansion (NTE) luminescence materials have triggered much concern due to the resistance of volume expansion with increasing temperature, which is a distinct advantage to practical applications.[17] In particular, the energy collection by activators can be promoted by the reversible lattice shrinkage and deformation, resulting the thermal enhanced emissions.[17, 18] For instance, Zou et al. reported NTE up-conversion $Yb_2W_3O_{12}$:Er crystal with more than 12 times luminescence enhancement when elevating temperature to 573 K.[18] Besides, Zhou et al. demonstrated that the NTE effect of $Zr(WO_4)_2$:Eu phosphors can effectively inhibit the emission loss of thermal quenching. Compared with room temperature, the luminescence intensity can be enhanced by 130% at 373 K.[19] Furthermore, the investigation on luminescence properties of lanthanide ions doped NTE $Sc_2W_3O_{12}$ material was carried out by researchers. Li et al reported $Eu^{3+}$-doped $Sc_2W_3O_{12}$ phosphors with good thermal quenching resistance at low temperature (97-280 K).[20] Recently, Wang et al. demonstrated that the emission intensity of $Sc_2W_3O_{12}$:$Eu^{3+}$ (x = 0.01–0.10) are abnormally enhanced with increasing temperatures up to 473 K.[21]

In this work, we have successfully synthesized $Sc_2W_3O_{12}$ phosphors with high $Eu^{3+}$ concentration (6-34 mol%). The luminescent thermal properties of $Sc_2W_3O_{12}$: $Eu^{3+}$ with various concentrations and their application in the field of optical temperature measurement were systematically investigated. Under the excitation at 264 nm, the bright red emission of $Sc_2W_3O_{12}$: $Eu^{3+}$ phosphors exhibit good thermally enhanced luminescence from 373 to 548 K. Specifically, the $Sc_2W_3O_{12}$: 6 mol% $Eu^{3+}$ phosphors presents 147.81% of initial intensity at 473 K. Besides, owing to the different responses of emissions at 592 nm ($^5D_0 \rightarrow {^7F_1}$) and 613 nm ($^5D_0 \rightarrow {^7F_2}$) to temperature, $Sc_2W_3O_{12}$: $Eu^{3+}$ phosphors show good temperature sensing properties based on FIR technique. The estimated maximum $S_a$ and $S_r$ are $6.872 \times 10^{-3}$ $K^{-1}$ and 3.063% $K^{-1}$ at 298 K, respectively. These results verify the availability of lanthanide ions doped in NTE luminescence materials against thermal quenching, and demonstrate the application in optical thermometers.

## 2. Experimental

### 2.1. Sample preparation

$Sc_2W_3O_{12}$ phosphors were prepared by high-temperature solid-state reaction. $Sc_2O_3$ (Scandium oxide, 99.99%, Aladdin), $WO_3$ (Tungsten trioxide, 99.99%, Aladdin) and $Eu_2O_3$ (Europium oxide, 99.99%, Aladdin) were used as raw materials. After weighed according to stoichiometry ratio, the raw materials were stirred in planetary ball mill for 12 h. After drying the obtained slurry, the mixture was pressed into thin sheets with a diameter of 15 mm and a thickness of 2 mm, and sintered in a box furnace at 1100 ℃ for 4 h. The sintered samples can be used for subsequent tests after being carefully ground in an agate mortar.

### 2.2. Characterization

The X-ray diffraction (XRD) of all samples were identified by the D8 ADVANCE X-ray diffractometer of the Bruker corporation with Cu-Kα radiation (λ=1.5406Å). Morphologies were characterized on a field-emission scanning electron microscope (FE-SEM, JSM-6700 F). Photoluminescence (PL) and PL excitation (PLE) spectra were measured using a fluorescence spectrophotometer (F-7000, Hitachi) equipped with a high-temperature fluorescence controller (TAP-02).

## 3. Results and discussion

### 3.1 Phase and morphology analysis

Fig. 1(a) shows the XRD results of $Sc_2W_3O_{12}$: $x$ mol% $Eu^{3+}$ phosphors ($x$= 6, 10, 14, 18, 22, 26, 30, 34). The diffraction peaks of samples are well matched with the standard patterns of $Sc_2W_3O_{12}$ (PDF#89-4690), demonstrating the successful synthesis of $Sc_2W_3O_{12}$ pure phase.[22] Fig. 1(b) depicts the schematic of the crystal structure of $Sc_2W_3O_{12}$, consisting of tetrahedral $WO_4$ and octahedral $ScO_6$ units connected by oxygen. The Rietveld refinement results indicate the orthorhombic structure with space group of Pnca (No. 60) of $Sc_2W_3O_{12}$: $Eu^{3+}$ phosphors at room temperature, as shown in Figs. 1(c) and S1. According to the refinement, we calculated the lattice parameters and corresponding volumes of various $Eu^{3+}$-doped $Sc_2W_3O_{12}$ phosphors and presented in Figs. 1 (d) and S2, respectively. Obviously, the volume of the crystal gradually increases with increasing the doping concentration of $Eu^{3+}$ ions. This can be attributed to that $Eu^{3+}$ (0.95 Å) with larger radius occupies the position of $Sc^{3+}$ (0.885 Å).[23]

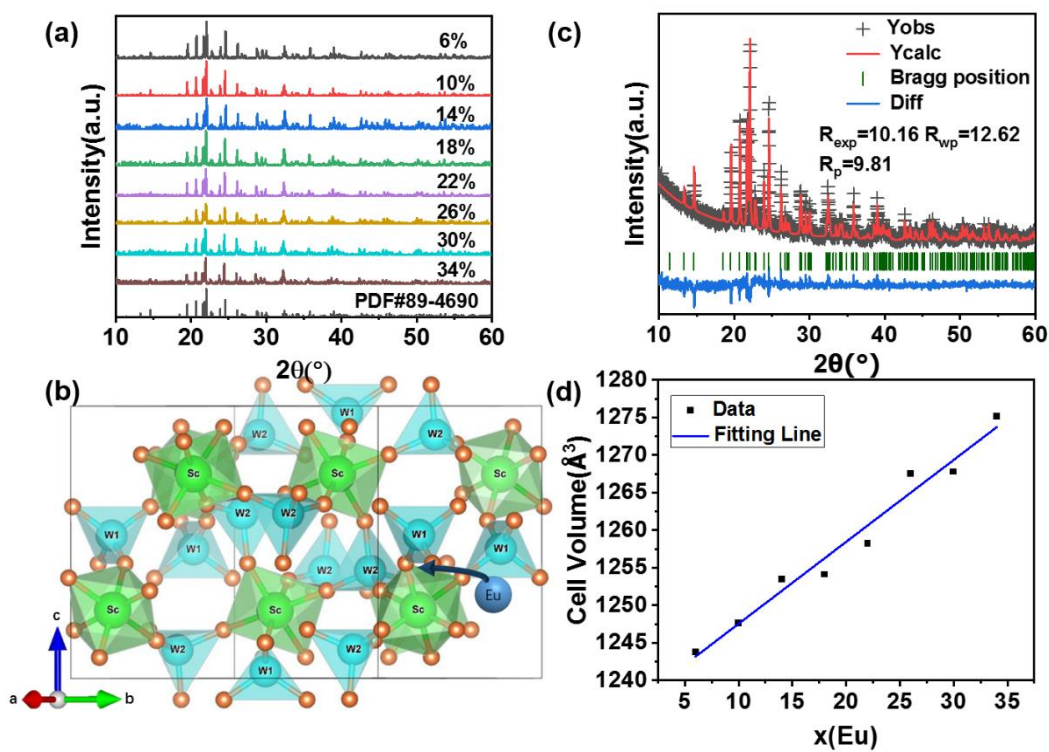

Fig.1 (a) the XRD patterns of $Sc_2W_3O_{12}$: $x$ mol% $Eu^{3+}$ ($x$= 6, 10, 14, 18, 22, 26, 30, 34) at room temperature. (b) Schematic of the crystal structure of orthorhombic $Sc_2W_3O_{12}$ and $Eu^{3+}$ ions substitution diagram. (c) Results of the Rietveld refinement of the XRD pattern of $Sc_2W_3O_{12}$: 6 mol% $Eu^{3+}$ phosphors. (d) Cell volume of $Sc_2W_3O_{12}$ with various $Eu^{3+}$ doping concentration.

The morphologies are recorded by SEM images in Fig. 2(a). The average size of $Sc_2W_3O_{12}$: $Eu^{3+}$ particles is 1 μm to 5 μm. Fig. 2(c-f) shows the element mappings of Sc, W, O and $Eu^{3+}$ in $Sc_2W_3O_{12}$:22 mol% $Eu^{3+}$ phosphors snatched in Fig. 2(b), respectively. It can be observed that the elements are uniformly distributed in $Sc_2W_3O_{12}$: $Eu^{3+}$ phosphors.

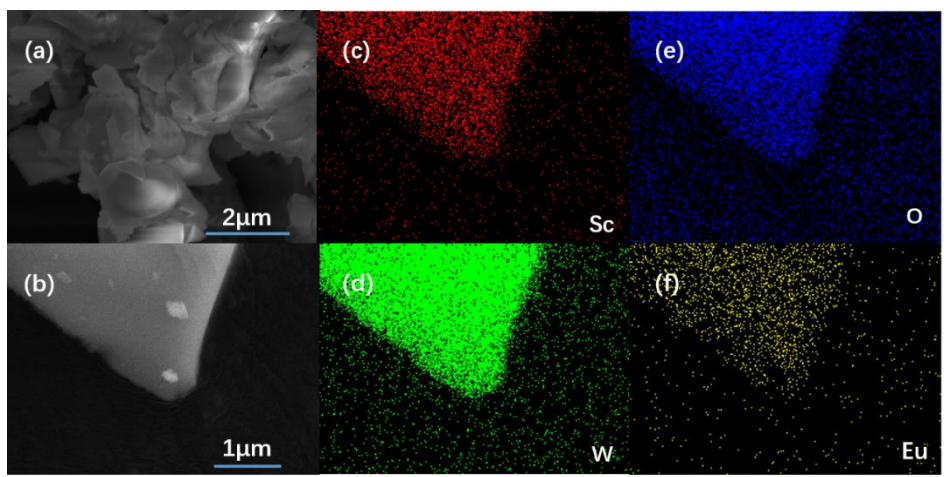

Fig.2 (a) and (b) are SEM images of $Sc_2W_3O_{12}$:22 mol% $Eu^{3+}$ phosphors. (c-f) The element mapping images of Sc, W, O and Eu snatched in (b), respectively.

### 3.2 Photoluminescence (PL) properties at room temperature

In order to study the luminescence properties of $Sc_2W_3O_{12}$: $Eu^{3+}$ phosphors, we measured the PL excitation (PLE) spectra of samples with different doping concentration of $Eu^{3+}$ ions at room temperature. The measurements were monitored at 613 nm emission, and the results are presented in Fig. 3(a). In general, the excitation spectrum of samples can be mainly divided into two parts. The broadband in the ultraviolet region (200-350 nm) derive from the charge transfer band (CTB) of tungstate ions.[19, 20] And the other sharp weak peaks at 318 nm, 360 nm, 384 nm, 393 nm and 464 nm are owing to transitions $^7F_0 \rightarrow {}^5D_4$, $^7F_0 \rightarrow {}^5L_7$, $^7F_0 \rightarrow {}^5L_6$, $^7F_0 \rightarrow {}^5D_3$ and $^7F_0 \rightarrow {}^5D_2$ of $Eu^{3+}$ ions, respectively. Fig. 3(b) illustrates the emission peaks of samples with different $Eu^{3+}$ concentrations under 264 nm excitation. And the peaks located at 592 nm, 613 nm and 653 nm result from the $^5D_0 \rightarrow {}^7F_1$, $^5D_0 \rightarrow {}^7F_2$ and $^5D_0 \rightarrow {}^7F_3$ transitions of $Eu^{3+}$ ions respectively. With increasing $Eu^{3+}$ concentration, the emission intensities initially increase and reach a maximum when x = 22, and then decrease due to the concentration quenching effect. The emission intensities of 592 nm and 613 nm are normalized to the emission of $Sc_2W_3O_{12}$: 22 mol% $Eu^{3+}$ phosphors, and their variation with $Eu^{3+}$ concentrations is depicted in Fig. 3(c). It can be observed that the emission intensity of 592 nm varying with $Eu^{3+}$ concentration is basically consistent with that of 613 nm. To explore the interaction type of the activators in $Sc_2W_3O_{12}$: $Eu^{3+}$ phosphors, Dexter theory is employed in Fig. 3(d) [24-26].

$$\ln\left(\frac{I}{x}\right) = A - \frac{\theta}{3}\ln(x)$$

where I is the integral fluorescence intensity, $x$ represents the doping concentration of $Eu^{3+}$, A is a constant, and θ refers to the eigenvalue related to the interaction type (θ =3, 6, 8, 10 corresponding to Exchange, dipole-dipole, dipole-quadrupole, quadrupole-quadrupole interactions, respectively)[26]. According to the fitting result, the θ value is about 3.0199, suggesting concentration quenching of activators in $Sc_2W_3O_{12}$: $Eu^{3+}$ phosphors derived from exchange interaction.

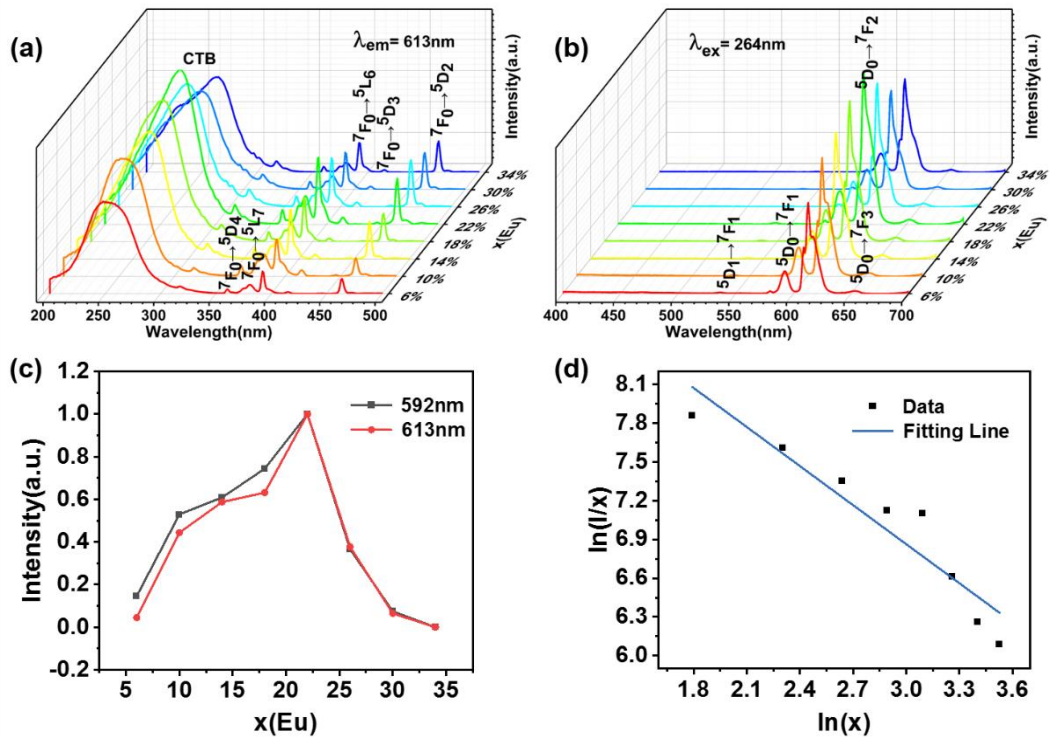

Fig.3 (a) PLE (emission at 613 nm) and (b) PL (excited by 264 nm) spectra of $Sc_2W_3O_{12}$: $x$ mol% $Eu^{3+}$ ($x$=6, 10, 14, 18, 22, 26, 30, 34) phosphors at room temperature. (c) Curves of normalized luminescence intensity at 592 and 613 nm varying with doping concentration. (d) Plot of In(I/$x$) vs In$x$, and the slope of the fitting line is 1.0065.

### 3.3 Temperature-dependent PL performances

Taking the $Sc_2W_3O_{12}$: 6 mol% $Eu^{3+}$ sample as an example to demonstrate the negative thermal expansion behavior of the $Sc_2W_3O_{12}$ matrix. The Rietveld refinement results (Fig. S2) of variable temperature XRD patterns presented in Fig. 4(a) verify the NTE of sample, and the refined lattice parameters as well as the cell volume values are illustrated in Fig. 4(b) and (c), respectively. Notably, the in-situ XRD patterns demonstrate the stable orthorhombic phase of $Sc_2W_3O_{12}$ at variable temperature. And the lattice of the sample shrinks in the direction of a-axis and c-axis, and expands slightly in the direction of b-axis from room temperature to 573 K. Consequently, the cell volume expands until 373 K and further contracts with rising temperature, which indicates the NTE of $Sc_2W_3O_{12}$: 6 mol% $Eu^{3+}$ phosphors ranging from 373 K to 573 K. As is known to all, changes in the crystal structure of materials will inevitably affect its luminescent properties. The temperature dependent emission spectra of $Sc_2W_3O_{12}$: 6 mol% $Eu^{3+}$ phosphors under the excitation at 264 nm is delineated in Fig. 4(d). Fig. 4(e) shows that the emission intensity at 613 nm maintains 86.61% of initial intensity at 373 K, then abnormally rapidly enhanced with increasing temperatures up to 548 K and the highest emission intensity reaches at 147.81% of initial intensity at 473 K. Moreover, the emissions at 592 nm present similar trends with 613 nm. It is noteworthy that the attenuate luminescence intensity from RT to 373 K can be ascribed to the thermal quenching, and the enhanced luminescence is due to the NTE effect from 373 to 548 K. In general, the bridging oxygen of Sc-O-W undergo transverse vibrations, leading to the coupled tilting of framework polyhedra.[22] As a result, the crystal structure of $Sc_2W_3O_{12}$: $Eu^{3+}$ phosphors grows denser and the energy collection of activator is promoted, demonstrating the stronger thermal quenching resistance of luminescence.[18, 21, 27] In addition, we measured the PL spectra of $Sc_2W_3O_{12}$: $x$ mol% $Eu^{3+}$ ($x$=10, 14, 18, 22, 26, 30) phosphors (Fig. S4). As shown in Fig. 4(f), it is obvious that the thermal quenching resistance of $Sc_2W_3O_{12}$: $Eu^{3+}$ decreases with the increase of $Eu^{3+}$ concentration at 298-548 K. At low $Eu^{3+}$ concentration (6-10 mol%), the PL intensity of the sample can maintain about 85% of initial intensity (in the green dashed box of Fig. 4(f)). However, at medium $Eu^{3+}$ concentration (14-18 mol%, in the orange dashed box) and high doping concentration (26-30 mol%, in the blue dashed box), the PL intensity of the sample can be reduced to about 80% and 67% of initial intensity, respectively. On the other hand, the onset temperature at which luminescence enhancement begins to appear also tends to increase with the increase of $Eu^{3+}$ doping concentration.

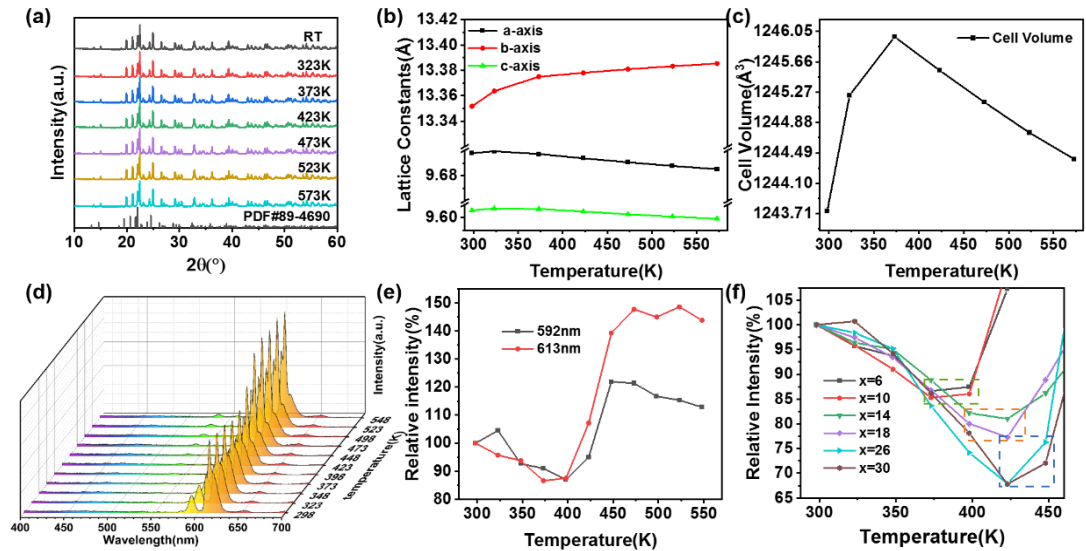

Fig.4 (a) Temperature dependent XRD patterns of $Sc_2W_3O_{12}$: 6 mol% $Eu^{3+}$ phosphors. (b) The lattice constants and (c) cell volume values of sample changing with temperature. (d) Temperature dependent PL spectra of $Sc_2W_3O_{12}$: 6 mol% $Eu^{3+}$ with 264 nm excitation. (e) Relative intensity at 592 nm and 613 nm emissions of $Sc_2W_3O_{12}$: 6 at% $Eu^{3+}$ varying with temperature. (f) Relative Intensity at 613 nm of $Sc_2W_3O_{12}$: $x$ mol% $Eu^{3+}$ (x=6, 10, 14, 18, 22, 26, 30) varying with temperature.

To investigate the temperature sensing properties of $Sc_2W_3O_{12}$: $x$ mol% $Eu^{3+}$ phosphors, the TCELs emissions at 592 nm and 613 nm are contributed to FIR technique. The FIR value R is defined as $R=I_{592\,nm}/I_{613\,nm}$, and depicted in Fig. 5(a) and Fig. S5. After fitted by a single-exponential function, the calculation of absolute sensitivity and relative sensitivity are executed. Generally, these two indicators are employed to evaluate the temperature sensing properties of luminescence materials. They can be estimated by equations.[30, 31]

$$S_a = \frac{dR(T)}{dT}$$

$$S_r = \frac{1}{R}\frac{dR(T)}{dT}$$

The calculated sensitivities are shown in Fig. 5(b) and Fig. S5. Comparing the maximum $S_a$ and $S_r$ values of samples under different $Eu^{3+}$ concentrations in Fig.5(c), we can see that the sensitivity of the sample decreases with the increase of $Eu^{3+}$ concentration. When the concentration reaches 22%, the maximum sensitivity begins to increase again. Among all samples, the sensitivity of the sample with 6 mol% $Eu^{3+}$ is the highest, and its maximum $S_a$ and $S_r$ values can reach $6.872\times10^{-3}$ $K^{-1}$ and 3.063% $K^{-1}$ at 298 K, respectively. In addition, the $S_a$ and $S_r$ values maintain the downward trends with increasing temperature, which implies possible preferable temperature sensing properties in low-temperature region.

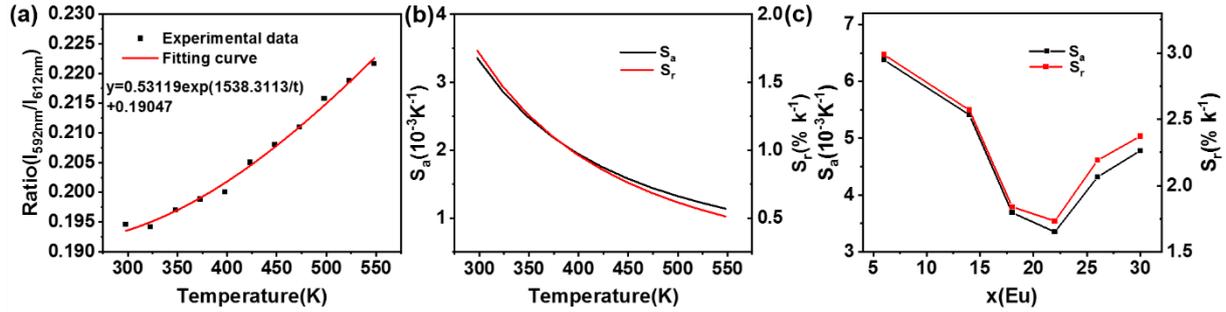

Fig.5 Temperature dependent (a) PL intensity ratio of 593nm and 613nm with fitting results and (b) absolute sensitivity ($S_a$) and relative sensitivity ($S_r$) values of $Sc_2W_3O_{12}$: 22 mol% $Eu^{3+}$ phosphors, The maximum $S_a$ and $S_r$ values varying with doping concentration.

## 4. Conclusion

In summary, we synthesized negative thermal expansion $Sc_2W_3O_{12}$:$Eu^{3+}$ phosphors by conventional solid-state method. XRD results reveal that the pure orthorhombic phase can be maintained when doping with various concentration of $Eu^{3+}$ ions or elevating temperature. Furthermore, the SEM images and elemental mappings confirm the uniform distribution of Sc, W, O and Eu ions in samples. Under 264 nm irradiation, the bright red emission can be observed and the quenching concentration of $Eu^{3+}$ ions is 22 mol% due to the exchange interaction. When increasing temperature from 298 K, the bridging oxygen of Sc-O-W undergo transverse vibrations, leading to the coupled tilting of framework polyhedra. And the lattices of $Sc_2W_3O_{12}$:$Eu^{3+}$ phosphors expand until 373 K and then contract up to 548 K. The emission intensity present abnormally rapidly enhanced with increasing temperatures derived from the enhanced energy collection of activators by lattice shrinkage and deformation. The ability of anti-thermal quenching decreases with increasing $Eu^{3+}$ doping concentration. In addition, resulting from the different responses of emissions at 592 nm ($^5D_0\rightarrow{}^7F_1$) and 613 nm ($^5D_0\rightarrow{}^7F_2$) to temperature, $Sc_2W_3O_{12}$: $Eu^{3+}$ phosphors can be used as optical thermometers based on FIR technique. The estimated maximum $S_a$ and $S_r$ values of $Sc_2W_3O_{12}$: 6 mol% $Eu^{3+}$ phosphors are $6.872\times10^{-3}$ $K^{-1}$ and 3.063% $K^{-1}$ at 298 K, respectively, These results demonstrate that the $Sc_2W_3O_{12}$: $Eu^{3+}$ phosphors can be applies as optical thermometers due to the good temperature sensing properties and thermal enhanced luminescence performance.

**Acknowledgements**

The authors thank the financial support of National Natural Science Foundation of China (No. 51872327).

# Supporting Information

# Thermally enhanced photoluminescence and temperature sensing properties of $Sc_2W_3O_{12}$:$Eu^{3+}$ phosphors

Yude Niu, Yuzhen Wang, Kaiming Zhu, Wanggui Ye, Zhe Feng, Hui Liu, Xin Yi, Yihuan Wang, Xuanyi Yuan[*]

Beijing Key Laboratory of Optoelectronic Functional Materials & Micro-nano Devices, Department of Physics, Renmin University of China, Beijing 100872, China.

Corresponding author:

Xuanyi Yuan: E-mail address: yuanxuanyi@ruc.edu.cn; telephone number: 0086-10-82501673; fax number: 0086-10-62517887; postal address: Renmin University of China, No. 59 Zhongguancun Street, Haidian District, Beijing 100872 P.R.China

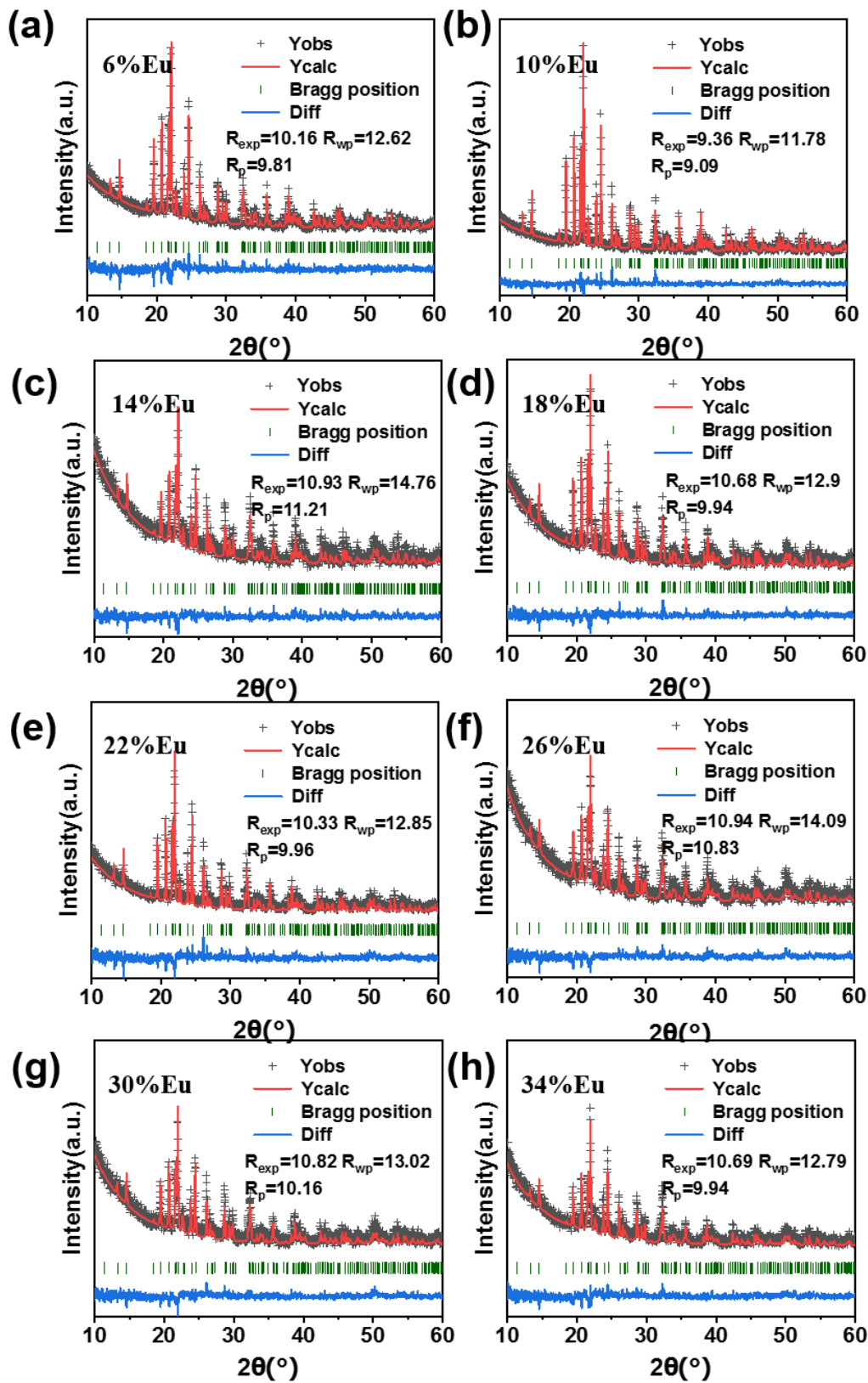

Fig.S1 (a-h)Results of the Rietveld refinement of the XRD pattern of $Sc_2W_3O_{12}$ : $x$ mol% $Eu^{3+}$ ($x$= 6, 10, 14, 18, 22, 26, 30, 34 ) at RT.

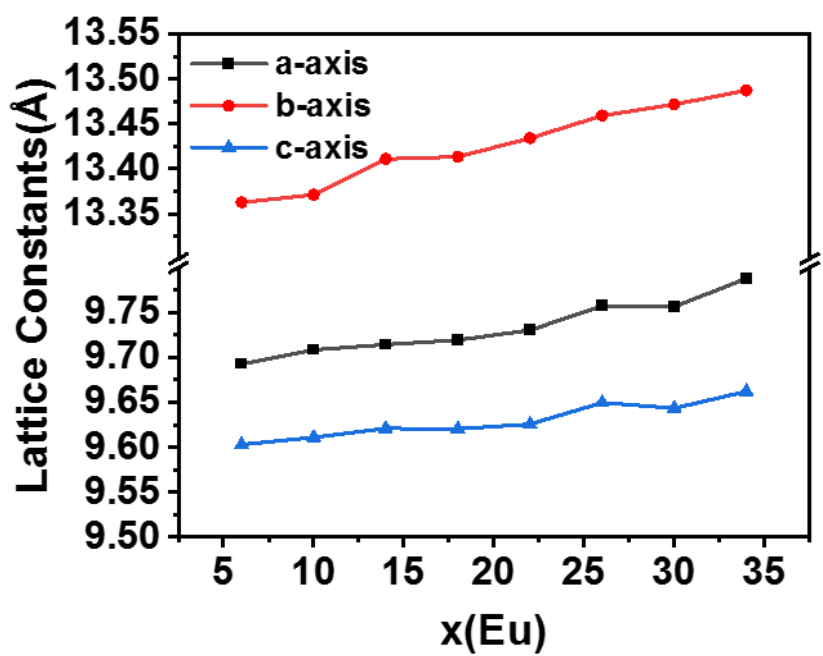

Fig.S2 Lattice parameters of $Sc_2W_3O_{12}$ changes with the doping concentration of $Eu^{3+}$ ions.

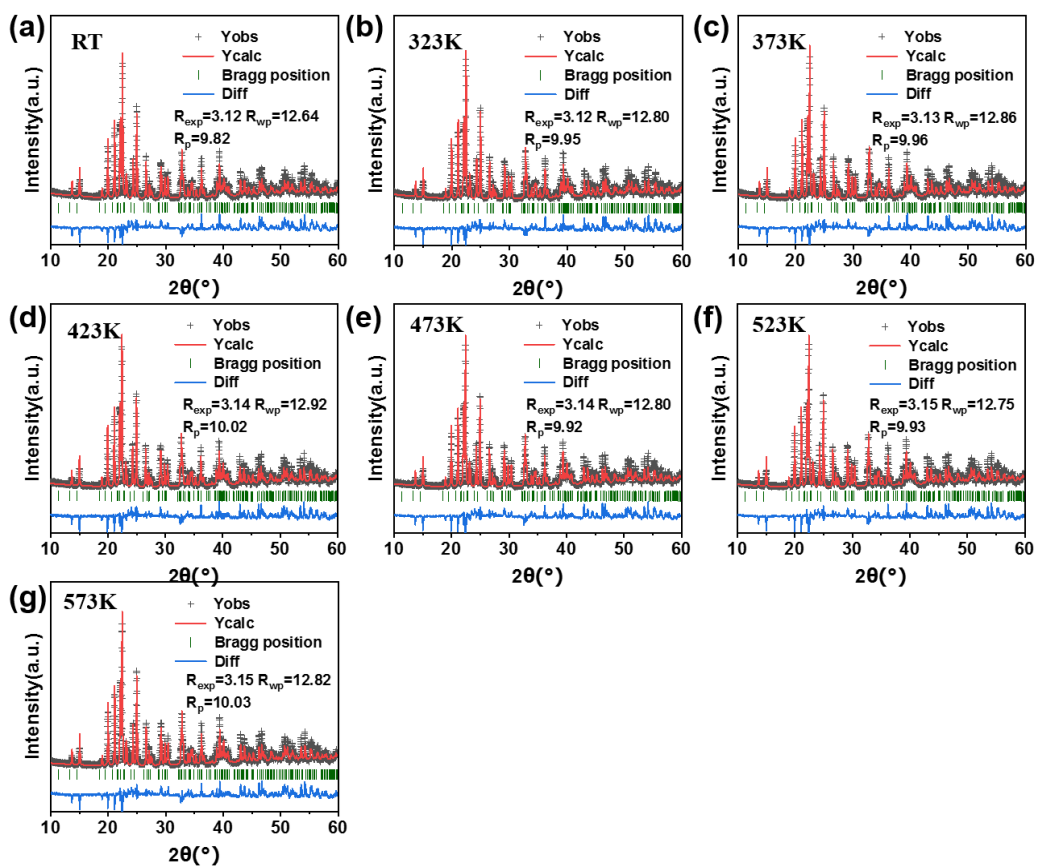

Fig.S3 Results of the Rietveld refinement of the XRD pattern of $Sc_2W_3O_{12}$ : 6 mol% $Eu^{3+}$ at different temperatures.

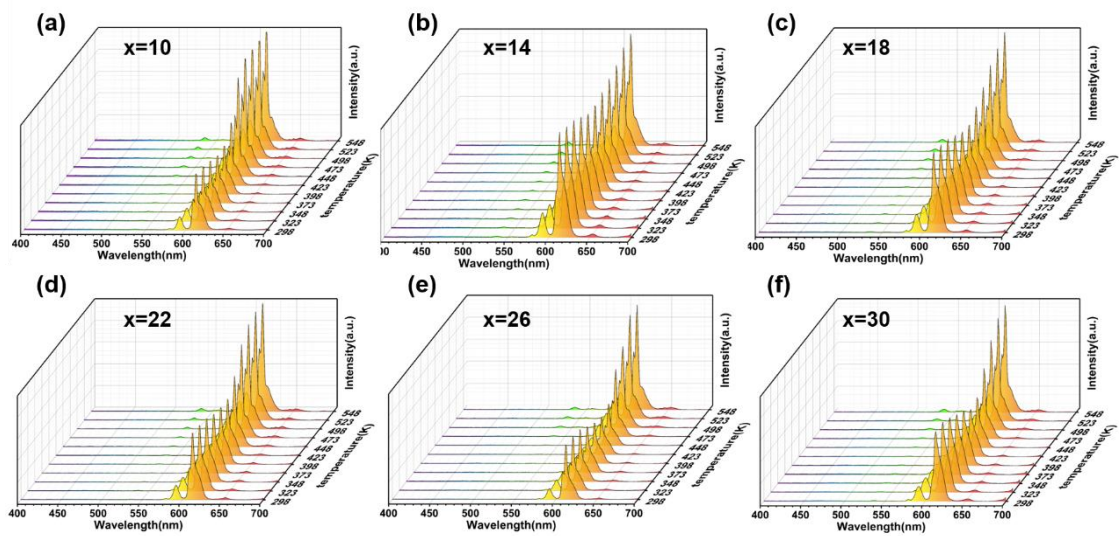

Fig.S4 (a-f) Temperature dependent XRD patterns of $Sc_2W_3O_{12}$: $x$ at% $Eu^{3+}$ ($x$=10, 14, 18, 22, 26, 30) phosphors.

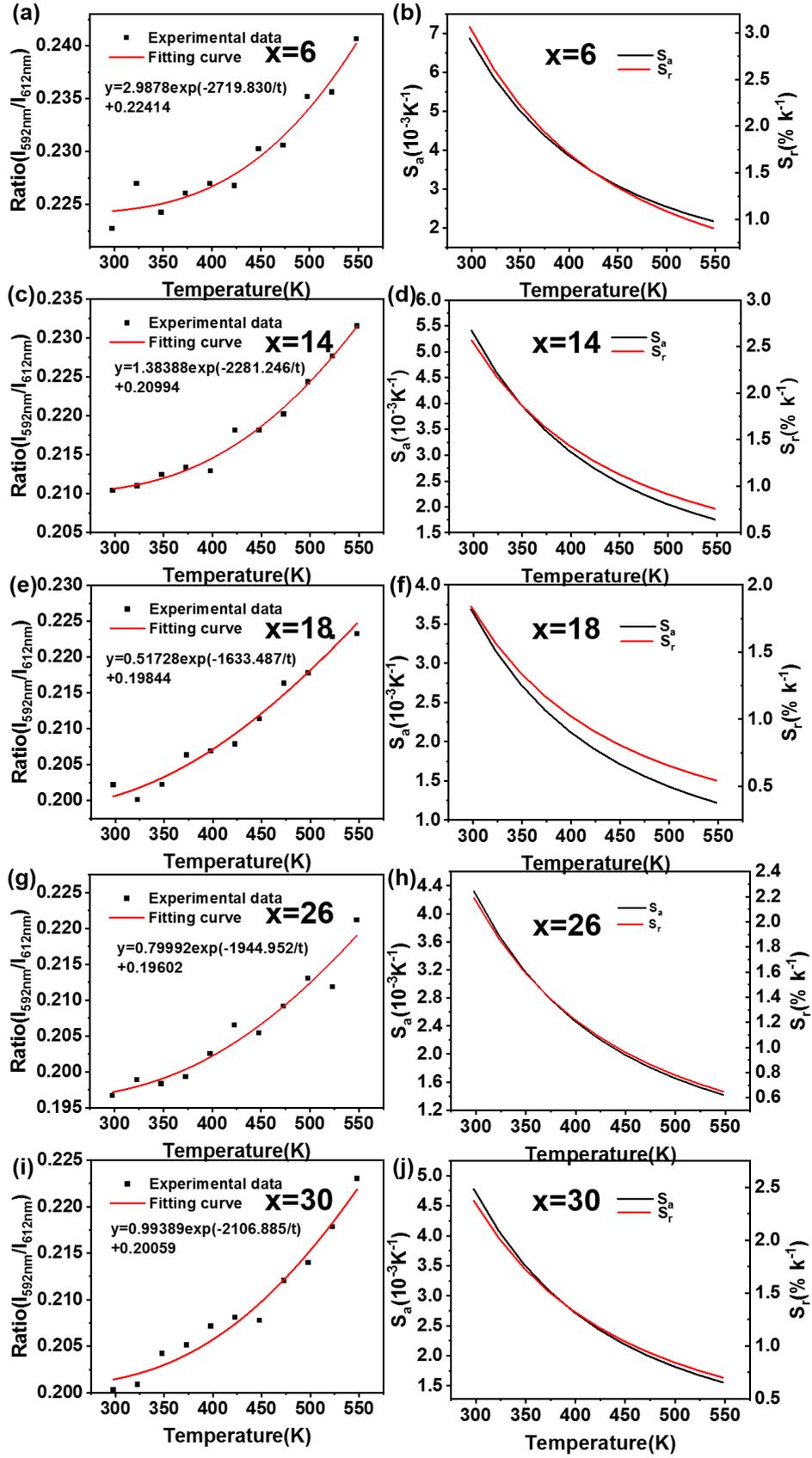

Fig.S5 Temperature dependent PL intensity ratio of 593nm and 613nm with fitting results and sensitivity values of $Sc_2W_3O_{12}$: $x$ mol% $Eu^{3+}$ ($x$=6, 14, 18, 26, 30).